\documentclass[onecolumn,showpacs,aps,floatfix,pra,superscriptaddress,nofootinbib]{revtex4}
\usepackage{color}
\usepackage{graphicx}
\usepackage{bm}
\usepackage{amsmath}
\begin{document}

\title{ Quantum effective force and Bohmian approach for time-dependent
traps}
\author{S. V. Mousavi}
\email{vmousavi@qom.ac.ir}
\affiliation{Department of Physics, The University of Qom, P. O. Box 37165, Qom, Iran}

\begin{abstract}

Trajectories of a Bohmian particle confined in time-dependent
cylindrical and spherical traps are computed for both contracting
and expanding boxes. Quantum effective force is considered in
arbitrary directions. It is seen that in contrast to the problem of
a particle in an infinite rectangular box with one wall in motion,
if particle initially is in an energy eigenstate of a tiny box the
force is zero in all directions. Trajectories of a two-body system
confined in the spherical trap are also computed for different
statistics.
Computations show that there are situations for which
distance between bosons is greater than the fermions. However,
results of average separation of the particles confirm
our expectation about the statistics.

\end{abstract}
\maketitle

\section{Introduction}

In real quantum world Hamiltonian of most systems is time-dependent
and thus solving the time dependent Schr\"{o}dinger equation, in
non-relativistic domain, is a needed one. A class of such systems is
systems with moving boundaries. Problem of a particle in a
one-dimensional infinite square-well potential with one wall in
uniform motion has been noticed from different aspects \cite{1D_Box,
Mo-PS-2012}. The concept of quantum effective force, time-derivative
of expectation value of the momentum operator, was introduced in
\cite{DoAn-PLA-2000-2002} in the context of the quantum deflection
of ultracold particles from mirrors. For the problem of a particle
in a 1D expanding infinite well potential it was shown that this
force apart from a minus sign is equal to the expectation value of
the gradient of the quantum potential \cite{Mo-PS-2012}.

Exact solution of the Schr\"{o}dinger equation has been found and
examined for a particle in a circular trap with a wall in uniform
motion \cite{Yu-PLA-2004} and for a particle in a hard sphere whose
wall is moving with a constant velocity \cite{Mo-EPL-2012}.

The aim of the present work is to probe some aspects of the
time-dependent boundary condition for a particle confined in a
circular and spherical box focusing on Bohmian interpretation of
quantum mechanics that have remained already unnoticed.
Although Bohmian formalism does not give predictions going beyond those of
quantum mechanics whenever the predictions of the later are
unambiguous, it should be used due to its interpretational
advantages.

The plan of the paper is as follows. In section \ref{sec: qmforce}
we consider quantum effective force in arbitrary directions for the
problem of a particle confined in time-dependent cylindrical and
spherical potential separately. In section \ref{sec: BM} we present
essentials of Bohmian trajectory approach to quantum mechanics; and
give the solution to the guidance equation when the particle is
initially in an energy eigenstate. This section contains numerical
calculations. Section \ref{sec: summary} gives the concluding
remarks.

\section{Quantum effective force} \label{sec: qmforce}

Time-derivative of the expectation value of the momentum operator of
a free particle in a box with volume $V$ bounded by the
moving closed surface $\sigma$ along the
arbitrary {\it fixed} direction $\hat{q}_0$ is given by
\begin{eqnarray}  \label{eq: 1}
\frac{d}{dt} \langle p_{q_{_0}} \rangle (t) &=& \frac{d}{dt} \int_V
dv~\psi^*({\bf{r}}, t) ~ \frac{\hbar}{i} \nabla_{q_{_0}} ~
\psi({\bf{r}}, t)
\nonumber \\
&=& \oint_{\sigma} \left( \psi^* \frac{\hbar}{i} \nabla_{q_{_0}}
\psi \right) {\bf{u}} . {\bf{da}} + \int_V dv \left[ \left(-i\hbar
\frac{\partial \psi^*}{\partial t} \right)  \nabla_{q_{_0}} \psi -
\psi^* \nabla_{q_{_0}}   \left( i\hbar \frac{\partial \psi}{\partial
t} \right) \right]~,
\end{eqnarray}
where we have used Leibnitz integral rule in the second equality;
and ${\bf{u}}$ shows the velocity of the surface $\sigma$. Now,
using the divergence theorem and the Schr\"{o}dinger equation
\begin{eqnarray} \label{eq: sch}
i \hbar \frac{\partial}{\partial t} \psi({\bf{r}}, t) &=&
-\frac{\hbar^2}{2 \mu} \nabla^2 \psi({\bf{r}}, t)~,
\end{eqnarray}
for the free particle we get
\begin{eqnarray} \label{eq: qmforce}
\frac{d}{dt} \langle p_{q_{_0}} \rangle (t) &=&
-\frac{\hbar^2}{2\mu} \oint_{\sigma} da ~ [( \hat{n}.
{\vec{\nabla}}) \psi^* ] \nabla_{q_{_0}} \psi~,
\end{eqnarray}
where we have used the vanishing on the surface boundary condition. $\hat{n}.{\vec{\nabla}}$
represents the normal component of the operator $\vec{\nabla}$.
%

%
%


\subsection{Quantum particle in an infinite cylindrical-well potential and quantum effective force}

Consider a particle with mass $\mu$ inside an infinite
cylindrical-well potential. The cylinder has a time-dependent radius
$L(t) = a + ut$ where $u$ is a constant; and a height $Z$, the
bottom and the top surfaces being at $z=0$ and $z=Z$. The exact
solutions of the Schr\"odinger equation (\ref{eq: sch}) for this
problem are given by \cite{Yu-PLA-2004}
\begin{eqnarray} \label{eq: norm_waves_cylind}
\psi_{m n k}(\rho, \phi, z, t) &=& \exp \left[ i \alpha \xi(t)
\left(\frac{\rho}{L(t)} \right)^2 - i x^2_{m n} \frac{1-1/\xi(t)}{4
\alpha} \right ] \frac{\sqrt{2}}{L(t)} \frac{1}{| J_{m + 1}(x_{m n})
|}  J_{m} \left( x_{m n}  \frac{\rho}{L(t)} \right) \frac{e^{ \pm
im\phi}}{\sqrt{2\pi}}
\nonumber \\
& \times &
\sqrt{\frac{2}{Z}} \sin{ \left( k \pi \frac{z}{Z} \right)
} \exp \left[ - i \frac{\hbar k^2 \pi^2}{2 \mu Z^2} t \right]
\nonumber\\
&\equiv&
f_{mn}(\rho, t) \frac{1}{\sqrt{\pi Z}}~e^{\pm im\phi} \sin{ \left( k \pi \frac{z}{Z} \right) }
\exp \left[ - i \frac{\hbar k^2 \pi^2}{2 \mu Z^2} t \right]
~,
\end{eqnarray}
where, $m = 0, 1, 2, ...$; $n = 1, 2, 3, ...$; $k = 1, 2, 3, ...$,
$J_{m}( x_{m n}) = 0$, $\alpha = \mu a u /(2 \hbar)$ and $\xi(t) =
L(t)/a$. The first line are solutions for a particle in a {\it
circular} box instead of a cylindrical one.
The general solution of eq. (\ref{eq: sch}) is a
superposition of functions (\ref{eq: norm_waves_cylind})
\begin{eqnarray} \label{eq: geral-sol_circular}
\psi({\bf{r}}, t) &=& \sum_{m^{\prime}=0}^{\infty}
\sum_{n^{\prime}=1}^{\infty} \sum_{k^{\prime}=1}^{\infty}
c_{m^{\prime} n^{\prime} k^{\prime}}\psi_{m^{\prime} n^{\prime}
k^{\prime}}({\bf{r}}, t)~,
\end{eqnarray}
with time-independent coefficients $c_{m^{\prime} n^{\prime}
k^{\prime}}$ determined from the relation
\begin{eqnarray} \label{eq: coef_circular}
c_{m^{\prime} n^{\prime} k^{\prime}} = \int d\rho ~ \rho \int d\phi
\int dz ~ \psi^*_{m^{\prime} n^{\prime} k^{\prime} }({\bf{r}}, 0)
\psi({\bf{r}}, 0) ~.
\end{eqnarray}
Eq. (\ref{eq: qmforce}) transforms to
\begin{eqnarray} \label{eq: force_cylinder}
\frac{d}{dt} \langle p_{q_{_0}} \rangle (t) &=&
-\frac{\hbar^2}{2\mu} \left[ ( - \int_{z=0} + \int_{z=Z} )
\rho~d\rho~d\phi \frac{\partial \psi^*}{\partial z} \nabla_{q_{_0}}
\psi +  \int_{\rho = L(t)} \rho~d\phi~dz \frac{\partial
\psi^*}{\partial \rho} \nabla_{q_{_0}} \psi \right],
\end{eqnarray}
in cylindrical coordinates $(\rho, \phi, z)$. From eqs. (\ref{eq:
norm_waves_cylind}) and (\ref{eq: geral-sol_circular}) one has,
\begin{eqnarray} \label{eq: psi_derivatives_cylinder}
\frac{\partial \psi}{\partial \rho} \bigg|_{z=0} = \frac{\partial
\psi}{\partial \rho} \bigg|_{z=Z} = \frac{\partial \psi}{\partial
\phi} \bigg|_{z=0}  = \frac{\partial \psi}{\partial \phi}
\bigg|_{z=Z} = \frac{\partial \psi}{\partial \phi}
\bigg|_{\rho=L(t)} = \frac{\partial \psi}{\partial z}
\bigg|_{\rho=L(t)} = 0 ~.
\end{eqnarray}
For arbitrary radial direction $\hat{\rho}_0 = \cos \phi_0 \hat{x} + \sin \phi_0 \hat{y}$ one obtains
\begin{eqnarray}
\nabla_{\rho_{_0}} \psi &=& \cos \phi_0 \frac{\partial
\psi}{\partial x} + \sin \phi_0 \frac{\partial \psi}{\partial y} =
\cos(\phi-\phi_0) \frac{\partial \psi}{\partial \rho} -
\sin(\phi-\phi_0) \frac{1}{\rho} \frac{\partial \psi}{\partial
\phi}~,
\end{eqnarray}
which from eq. (\ref{eq: psi_derivatives_cylinder}) becomes zero at both
the bottom and the top surfaces $z=0$ and $z=Z$; and the second term of the
right-hand-side is zero at the lateral surface $\rho = L(t)$.
Thus, we get
\begin{eqnarray} \label{eq: force_cylinder_rho}
\frac{d}{dt} \langle p_{\rho_{_0}} \rangle (t) &=&
-\frac{\hbar^2}{2\mu} L(t) \int_0^{2\pi} d\phi ~ \cos(\phi-\phi_0)
\int_0^Z dz \bigg| \frac{\partial \psi}{\partial \rho}(\rho = L(t),
\phi) \bigg|^2
\nonumber \\
&=& -\frac{\hbar^2}{4\mu} L(t) \sum_{m^{\prime} n^{\prime}}
\sum_{m n k}
c^*_{m^{\prime} n^{\prime} k} c_{mnk} \frac{\partial f^*_{m^{\prime} n^{\prime}}}{\partial \rho} \bigg|_{\rho=L(t)}
\frac{\partial f_{m n}}{\partial \rho}\bigg|_{\rho=L(t)}
( e^{-i\phi_0} \delta_{\mp m \pm m^{\prime}, 1} + e^{i\phi_0} \delta_{\pm m \mp m^{\prime}, 1} ) ~.
\end{eqnarray}
One can similarly obtain quantum effective force
\begin{eqnarray} \label{eq: force_cylinder_phi}
\frac{d}{dt} \langle p_{\phi_{_0}} \rangle (t) &=&
-\frac{\hbar^2}{2\mu} L(t) \int_0^{2\pi} d\phi ~ \sin(\phi-\phi_0)
\int_0^Z dz \bigg| \frac{\partial \psi}{\partial \rho}(\rho = L(t),
\phi) \bigg|^2
\nonumber \\
&=& -\frac{\hbar^2}{4 \mu i} L(t) \sum_{m^{\prime} n^{\prime}}
\sum_{m n k}
c^*_{m^{\prime} n^{\prime} k} c_{mnk} \frac{\partial f^*_{m^{\prime} n^{\prime}}}{\partial \rho} \bigg|_{\rho=L(t)}
\frac{\partial f_{m n}}{\partial \rho}\bigg|_{\rho=L(t)}
( e^{-i\phi_0} \delta_{\mp m \pm m^{\prime}, 1} - e^{i\phi_0}  \delta_{\pm m \mp m^{\prime}, 1} ) ~,
\end{eqnarray}
along arbitrary azimuthal direction $\hat{\phi}_0$ and
\begin{eqnarray} \label{eq: force_cylinder_z}
\frac{d}{dt} \langle p_{z} \rangle (t) &=& -\frac{\hbar^2}{2\mu} (
-\int_{z=0} + \int_{z=Z} ) \rho~d\rho~d\phi \bigg| \frac{\partial
\psi}{\partial z} \bigg|^2
\nonumber\\
&=&
 -\frac{\hbar^2 \pi^2}{\mu Z^3} \sum_{m, n, k, k^{\prime}} c_{mnk}~
 c^*_{mnk^{\prime}}~k~k^{\prime}~[ -1 + (-1)^{k+k^{\prime}} ]
\exp \left[ - i \frac{\hbar ( k^2 - {k^{\prime}}^2 ) \pi^2}{2 \mu Z^2} t \right]
~.
\end{eqnarray}
along direction $\hat{z}$.

When the initial wavefunction is an energy eigenstate of a smaller box
with radius $a_{\text{s}} < a$,
\begin{eqnarray}
\psi({\bf{r}}, 0) &=& \frac{\sqrt{2}}{a_{\text{s}}}  \frac{1}{| J_{m
+ 1}(x_{m n}) |} J_{m} \left( x_{m n}  \frac{\rho}{a_{\text{s}}}
\right) \frac{e^{im\phi}}{\sqrt{2\pi}} \sqrt{\frac{2}{Z}} \sin{
\left( k \pi \frac{z}{Z} \right) } ~,
\end{eqnarray}
then
\begin{eqnarray}  \label{eq: cir_special_sol}
\psi({\bf{r}}, t) &=& \sum_{n^{\prime}} I_{mnn^{\prime}} (\alpha)
\psi_{m n^{\prime} k}({\bf{r}}, t)
\end{eqnarray}
where,
\begin{eqnarray} \label{eq: I_int_circ}
I_{mnn^{\prime}} (\alpha) &=& \frac{2}{a_{\text{s}} a} \frac{1}{|
J_{m + 1}(x_{m n}) |} \frac{1}{| J_{m + 1}(x_{m n^{\prime}}) |}
 \int_0^{a_{\text{s}}} d\rho ~ \rho ~ e^{- i \alpha (\rho /a)^2 } J_{m} \left( x_{m n}  \frac{\rho}{a_{\text{s}}} \right)
J_{m} \left( x_{m n^{\prime}}  \frac{\rho}{a} \right)~.
\end{eqnarray}
In such a case the modulus of $\psi$ and its derivatives are
independent of azimuthal angle $\phi$ and thus integral over $\phi$
of (\ref{eq: force_cylinder_rho}), (\ref{eq: force_cylinder_phi})
and (\ref{eq: force_cylinder_z}) is zero. Therefore, quantum
effective force is zero in any arbitrary direction for the problem of
a particle in a cylindrical trap.

%
%
%
\subsection{Quantum particle in an infinite spherical-well potential and quantum effective force}

For a particle with mass $\mu$ inside an infinite spherical-well
potential with radius $L(t) = a + ut$, exact solutions of the
Schr\"{o}dinger equation (\ref{eq: sch}) read \cite{Mo-EPL-2012}
\begin{eqnarray} \label{eq: norm_waves_spherical}
\psi_{l n m}({\bf{r}}, t) &=& \exp \left[ i \alpha \xi(t)
\left(\frac{r}{L(t)} \right)^2 - i x^2_{l n} \frac{1-1/\xi(t)}{4
\alpha} \right] \sqrt{\frac{2}{L^3(t)}} \frac{1}{| j_{l + 1}(x_{l
n}) |} j_{l} \left( x_{l n}  \frac{r}{L(t)} \right)
 Y_{l m}(\theta, \phi) ~,
\nonumber\\
&\equiv&
g_{ln}(r, t)  Y_{l m}(\theta, \phi)~,
\end{eqnarray}
where, $l = 0, 1, 2, ...$; $m = -l, -l+1, ..., 0, ..., l-1, l$; $n = 1, 2, 3, ...$ and
$j_{l}( x_{l n}) = 0$.
The general solution of eq. (\ref{eq: sch}) is a
superposition of functions (\ref{eq: norm_waves_spherical})
\begin{eqnarray} \label{eq: geral-sol_spherical}
\psi({\bf{r}}, t) &=& \sum_{l^{\prime}=0}^{\infty}
\sum_{n^{\prime}=1}^{\infty} \sum_{m^{\prime} =
-l^{\prime}}^{l^{\prime}} c_{l^{\prime} n^{\prime}
m^{\prime}}\psi_{l^{\prime} n^{\prime} m^{\prime}}({\bf{r}}, t)~,
\end{eqnarray}
with time-independent coefficients $c_{l^{\prime} n^{\prime}
m^{\prime}}$ determined from the relation
\begin{eqnarray} \label{eq: coef_spherical}
c_{l^{\prime} n^{\prime} m^{\prime}} = \int_0^a dr ~ r^2 \int
d\Omega ~ \psi^*_{l^{\prime} n^{\prime} m^{\prime}}({\bf{r}}, 0)
\psi({\bf{r}}, 0) ~.
\end{eqnarray}
In spherical coordinates $(r, \theta, \phi)$ eq.(\ref{eq: qmforce})
transforms to
\begin{eqnarray} \label{force_sphere}
\frac{d}{dt} \langle p_{q_{_0}} \rangle (t) &=&
-\frac{\hbar^2}{2\mu} L^2(t) \int d\Omega (\frac{\partial
\psi^*}{\partial r} \nabla_{q_{_0}} \psi) \bigg|_{r=L(t)}~.
\end{eqnarray}
From eqs. (\ref{eq: norm_waves_spherical}) and (\ref{eq: geral-sol_spherical}) one obtains
\begin{eqnarray} \label{eq: psi_der_sphere}
\frac{\partial \psi}{\partial \theta} \bigg|_{r=L(t)} &=&
\frac{\partial \psi}{\partial \phi} \bigg|_{r=L(t)} = 0~.
\end{eqnarray}
For an arbitrary radial direction
$\hat{r}_0 = (\hat{x} \cos \phi_0 + \hat{y} \sin \phi_0) \sin \theta_0 + \hat{z} \cos \theta_0$, one has
\begin{eqnarray}
\nabla_{r_{_0}} \psi &=& \hat{r}_0.\vec{\nabla}\psi = [ \sin
\theta_0 \sin \theta \cos (\phi-\phi_0) + \cos \theta_0 \cos \theta]
\frac{\partial \psi}{\partial r}
\nonumber\\
&+& [ \sin \theta_0 \cos \theta \cos (\phi-\phi_0) - \cos \theta_0
\sin \theta] \frac{1}{r} \frac{\partial \psi}{\partial \theta} -\sin
\theta_0 \sin (\phi-\phi_0) \frac{1}{r \sin \theta} \frac{\partial
\psi}{\partial \phi}~,
\end{eqnarray}
and thus from (\ref{eq: psi_der_sphere})
\begin{eqnarray}
\nabla_{r_{_0}} \psi \bigg|_{r=L(t)} &=& [ \sin \theta_0 \sin \theta
\cos (\phi-\phi_0) + \cos \theta_0 \cos \theta] \frac{\partial
\psi}{\partial r} \bigg|_{r=L(t)}~.
\end{eqnarray}
Ultimately, we get
\begin{eqnarray} \label{eq: force_sphere_r}
\frac{d}{dt} \langle p_{r_{_0}} \rangle (t) &=& \frac{\hbar^2}{2\mu}
L^2(t) \sqrt{ \frac{4\pi}{3} } \int d\Omega ~\left[ \sqrt{2} \sin
\theta_0 ~{\text{Re}} \left( e^{-i\phi_0} Y_{11}(\Omega) \right) - \cos
\theta_0 Y_{10}(\Omega)\right] ~ \bigg| \frac{\partial
\psi}{\partial r} (r = L(t), \Omega) \bigg|^2 ~.
\end{eqnarray}
In the same way, one can show
\begin{eqnarray}
\frac{d}{dt} \langle p_{\theta_{_0}} \rangle (t) &=&
\frac{\hbar^2}{2\mu} L^2(t) \sqrt{ \frac{4\pi}{3} }
\int d\Omega ~ \left[\sqrt{2} \cos \theta_0 ~{\text{Re}} \left( e^{-i\phi_0} Y_{11}(\Omega) \right) +
\sin \theta_0 Y_{10}(\Omega) \right] ~ \bigg| \frac{\partial \psi}{\partial r} (r = L(t), \Omega) \bigg|^2
~,  \label{eq: force_sphere_theta}\\
\frac{d}{dt} \langle p_{\phi_{_0}} \rangle (t) &=&
\frac{\hbar^2}{2\mu} L^2(t) \sqrt{ \frac{8\pi}{3} }~ {\text{Im}} \left(
e^{-i\phi_0} \int d\Omega ~Y_{11}(\Omega) ~ \bigg| \frac{\partial
\psi}{\partial r} (r = L(t), \Omega) \bigg|^2 \right) ~, \label{eq:
force_sphere_phi}
\end{eqnarray}
where eqs. (\ref{eq: force_sphere_theta}) and (\ref{eq:
force_sphere_phi}) display respectively quantum effective force in
arbitrary directions $\hat{\theta}_0 = (\hat{x} \cos \phi_0 +
\hat{y} \sin \phi_0) \cos \theta_0 - \hat{z} \sin \theta_0$ and
$\hat{\phi}_0 = -\hat{x} \sin \phi_0 + \hat{y} \cos \phi_0$.

When the initial wavefunction is an energy eigenstate of a smaller
box with radius $a_{\text{s}} < a$,
\begin{eqnarray}
\psi({\bf{r}}, 0) &=& \sqrt{\frac{2}{a_{\text{s}}^3}}  \frac{1}{|
j_{l + 1}(x_{l n}) |} j_{l} \left( x_{l n}  \frac{r}{a_{\text{s}}}
\right)
 Y_{l m}(\theta, \phi) ~,
\end{eqnarray}
then
\begin{eqnarray} \label{eq: psi_eigen}
\psi({\bf{r}}, t) &=& \sum_{n^{\prime}} I_{lnn^{\prime}} (\alpha)
\psi_{l n^{\prime} m}({\bf{r}}, t)~,
\end{eqnarray}
where,
\begin{eqnarray} \label{eq: I_int_sphere}
I_{lnn^{\prime}} (\alpha) &=& {\frac{2}{\sqrt{a^3 a_{\text{s}}^3}}}
\frac{1}{| j_{l + 1}(x_{l n}) |} \frac{1}{| j_{l + 1}(x_{l
n^{\prime}}) |}
 \int_0^{a_{\text{s}}} dr ~ r^2 e^{- i \alpha (r/a)^2 } j_{l} \left( x_{l n}  \frac{r}{a_{\text{s}}} \right)
j_{l} \left( x_{l n^{\prime}}  \frac{r}{a} \right)~.
\end{eqnarray}
In such a case $|\partial \psi/\partial r|^2$ is independent of the
azimuthal angle $\phi$, thus the first integral of (\ref{eq:
force_sphere_r}) and (\ref{eq: force_sphere_theta}); and the
integral of (\ref{eq: force_sphere_phi}) would be zero.
$|Y_{lm}(\theta, \phi)|^2$ is an even function of $\cos \theta$ and
thus the second integral of (\ref{eq: force_sphere_r}) and (\ref{eq:
force_sphere_theta}) is also zero. Therefore, quantum effective
force is zero in any arbitrary direction while the force is not
zero for the corresponding problem in 1D \cite{Mo-PS-2012}.
%
%
%
\section{Bohmian trajectories} \label{sec: BM}

In causal interpretation of quantum mechanics \cite{BoI&II_PR_1952,
Ho_book_1993} by introducing polar form $\psi({\bf{r}}, t) = R
\exp{(iS/\hbar)}$ in the Schr\"{o}dinger equation (\ref{eq: sch})
one obtains
\begin{eqnarray} \label{eq: HJ}
\frac{\partial S}{\partial t} + \frac{(\vec{\nabla} S)^2}{2\mu} + V + Q &=& 0 ~,\\
\frac{\partial R^2}{\partial t} + \vec{\nabla} . \left( {R^2 \frac{\vec{\nabla} S}{\mu}} \right) &=& 0 ~,
\end{eqnarray}
where $Q({\bf{r}}, t) = (-\hbar^2/2 \mu) \nabla^2 R/R$ is called quantum potential energy.
In this theory particle trajectory ${\bf{r}}(t)$ is determined by simultaneous integration
of the time-dependent Schr\"{o}dinger equation and the guidance equation
\begin{eqnarray} \label{eq: guaidance}
\frac{ d{\bf{r}}(t) }{d t} &=& {\bf{v}}({\bf{r}}(t), t) =
\frac{\vec{\nabla} S}{\mu} \bigg|_{ {\bf{r}}(t) } =
\frac{\hbar}{\mu}~{\text{Im}} \left( {\frac{\vec{\nabla}\psi({\bf{r}},
t)}{\psi({\bf{r}}, t)}} \right) \bigg|_{ {\bf{r}} = {\bf{r}}(t) }~,
\end{eqnarray}
by specifying the initial condition ${\bf{r}}(0) = {\bf{r}}_0$.

In this context one can easily show that the
time-derivative of the expectation value of {\it actual} momentum of the
particle, ${\bf{p}} = \vec{\nabla} S$, is given by \cite{Mo-PS-2012}
\begin{eqnarray} \label{eq: BM_qmforce}
\frac{d}{dt} \langle \nabla_{q_{_0}} S \rangle (t) &=&
\int_V dv~ \left[ \frac{\partial R^2}{\partial t} \nabla_{q_{_0}} S + R^2  \nabla_{q_{_0}} \frac{\partial S}{\partial t} \right]
= - \int_V dv~ R^2 \nabla_{q_{_0}} Q = \langle - \nabla_{q_{_0}} Q \rangle (t)~.
\end{eqnarray}
Using the polar form of the wavefunction in the second line of eq. (\ref{eq: 1}) and vanishing on the surface boundary condition
we obtain the equality $\frac{d}{dt} \langle p_{q_{_0}} \rangle (t) = \frac{d}{dt} \langle \nabla_{q_{_0}} S \rangle (t)$.
Respecting to the Newton's second law $ \frac{d}{dt} {\bf{p}} = - \vec{\nabla} (Q+V) $ in Bohmian mechanics, last equality in eq . (\ref{eq: BM_qmforce}) shows the relation $ \langle \frac{d}{dt} {\bf{p}} \rangle = \frac{d}{dt} \langle {\bf{p}} \rangle $ for the actual momentum of the particle.
Although, the classical potential energy $V({\bf{r}}, t)$ is zero inside the box,
the quantum potential energy is non-zero. This is the reason of
interaction of the boundary with the confined particle.

In the following we will compute trajectories of a Bohmian particle
which is initially in an energy eigenstate of the box. So, $a_s$ in
eqs. (\ref{eq: I_int_circ}) and (\ref{eq: I_int_sphere}) is replaced
with $a$. For numerical calculations new quantities $v_{mn} = \hbar
x_{mn} / \mu a$ with dimension of velocity, $t_{mn} = a/v_{mn} = \mu
a^2 / \hbar x_{mn}$ with dimension of time and dimensionless
quantity $\alpha_{mn} = \mu a v_{mn}/2\hbar = x_{mn}/2$ are defined
for the problem of a particle in a circular trap. Corresponding
quantities in the case of a particle in a spherical trap are $v_{ln}
= \hbar x_{ln} / \mu a$, $\alpha_{ln} = x_{ln}/2$ and $t_{ln} = \mu
a^2 / \hbar x_{ln}$. In our calculations velocity of the moving wall
is determined versus the above velocities.
\subsection{circular box}

\begin{figure}
\centering
\includegraphics[width=12cm,angle=-90]{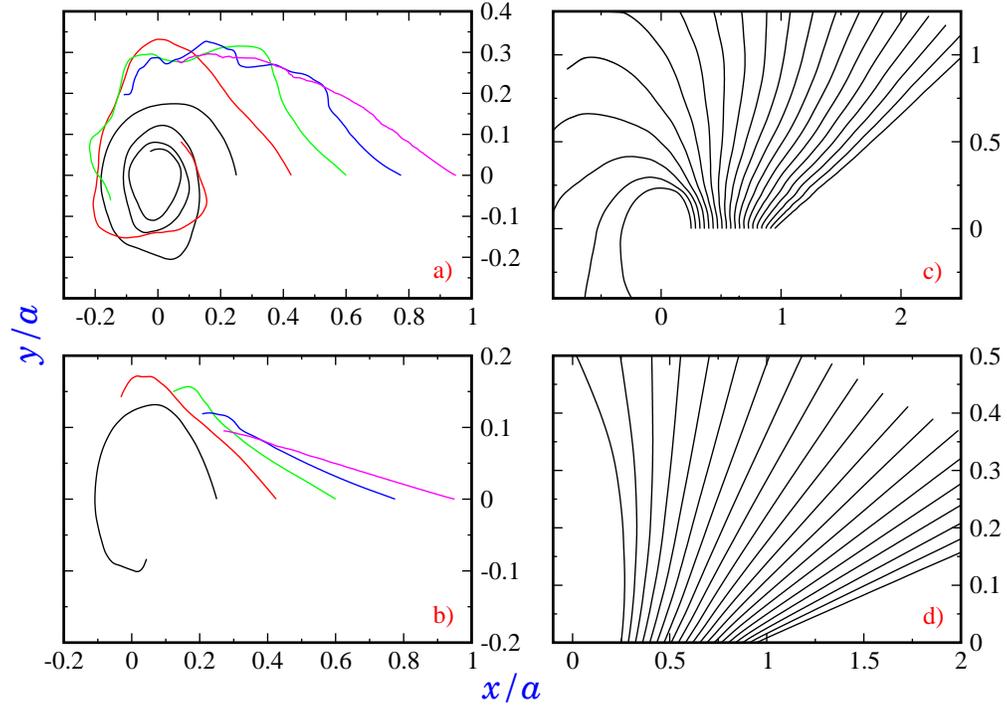}
\caption{(Color online) A selection of Bohmian trajectories for a
particle in a time dependent circular trap for different values of
wall velocity: a) $\alpha = -0.5 \alpha_{11}$, b) $\alpha = -2
\alpha_{11}$, c) $\alpha = 0.5 \alpha_{11}$ and d) $\alpha = 2
\alpha_{11}$. Particle is initially in the energy eigenstate
$u_{11}(\rho, \phi)$. }
\vspace*{0.1cm}
\label{fig: traj_circular}
\end{figure}

Using eqs. (\ref{eq: guaidance}) and (\ref{eq: cir_special_sol}) one finds,
\begin{eqnarray}
\dot{\rho} &=& \frac{\hbar}{\mu} ~{\text{Im}} \left( \frac{\sum_{n^{\prime}}
I_{m n n^{\prime}} (\alpha)~\frac{\partial f_{m
n^{\prime}}}{\partial \rho} }{\sum_{n^{\prime}} I_{m n n^{\prime}}
(\alpha) f_{m n^{\prime}} } \right)~, \label{eq: v_rho_circular}
\\
\dot{\phi} &=& \frac{\hbar}{\mu} \frac{m}{\rho^2} ~. \label{eq: v_phi_circular}
\end{eqnarray}
$\rho(t)$ is found by numerical solving of eq. (\ref{eq: v_rho_circular})
and then by using this result in (\ref{eq: v_phi_circular}) one obtains
\begin{eqnarray}
\phi(t) &=& \phi_0 + \frac{m \hbar}{\mu} \int_0^t \frac{dt^{\prime}}{\rho^2(t^{\prime})}
\end{eqnarray}
Trajectories for the case of a particle in circular box are
displayed in $xy$ plane in figure \ref{fig: traj_circular} for
different rates of contraction and expansion. Here, the particle is
initially in the energy eigenstate
$u_{11}(\rho, \phi) = \sqrt{2} J_1(x_{11} \rho /a) \exp{(i\phi)}/a |J_2(x_{11})|$.
%
%
It should be noted that trajectories cross each other at different times
(different values of the $\phi$-coordinate). Thus, there is no problem regarding
the single-valuedness of the wavefunction.
\subsection{spherical box}

Eqs. (\ref{eq: guaidance}) and (\ref{eq: I_int_sphere}) lead to
\begin{eqnarray}
\dot{r} &=& \frac{\hbar}{\mu} ~{\text{Im}} \left(\frac{\sum_{n^{\prime}}
I_{l n n^{\prime}} (\alpha) ~
\partial g_{l n^{\prime}} / \partial r }{\sum_{n^{\prime}} I_{l n
n^{\prime}} (\alpha)~ g_{l n^{\prime}} } \right)~, \label{eq:
v_r_spherical}
\\
\dot{\theta} &=& \frac{\hbar}{\mu} \frac{1}{r^2} ~ {\text{Im}} \left(
\frac{\partial Y_{l, m}(\theta, \phi) / \partial \theta}{Y_{l,
m}(\theta, \phi)} \right) = 0 ~, \label{eq: v_theta_spherical}
\\
\dot{\phi} &=& \frac{\hbar}{\mu} \frac{m}{r^2 \sin^2 \theta} ~, \label{eq: v_phi_spherical}
\end{eqnarray}
for the velocity of Bohmian particle in a spherical trap and
initially in an energy eigenstate.
From eq. (\ref{eq: v_theta_spherical}) one sees that during the
motion, polar angle $\theta$ will not change, i.e., motion would be
on the surface of a cone with opening angle $\theta = \theta_0$. By
numerical solution of eq. (\ref{eq: v_r_spherical}) one finds $r(t)$
and then eq. (\ref{eq: v_phi_spherical}) is solved,
\begin{eqnarray}
\phi(t) &=& \phi_0 + \frac{m \hbar}{\mu \sin ^2 \theta_0} \int_0^t
\frac{ dt^{\prime} }{r^2 (t^{\prime})}~.
\end{eqnarray}
From this equation it's apparent that for $m=0$ azimuthal angle
$\phi$ does not change during the motion which is an obvious result:
angular momentum in $\hat{z}$-direction is zero in such a case.
Thus, in this case motion is along the straight line $\phi = \phi_0$.

\begin{figure}
\centering
\includegraphics[width=12cm,angle=-90]{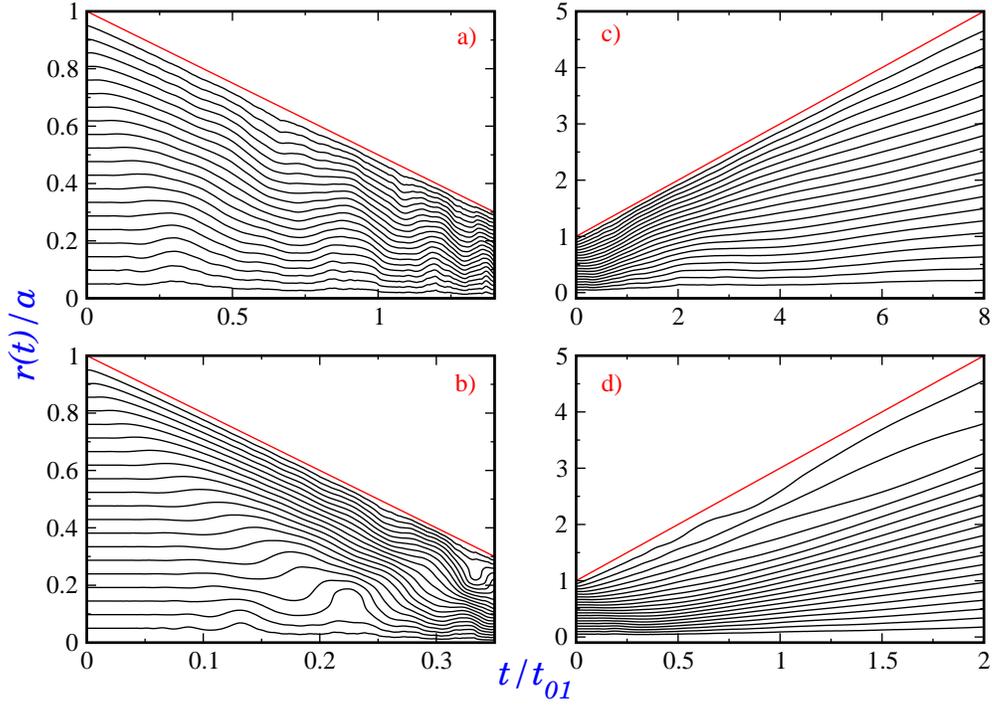}
\caption{(Color online) A selection of Bohmian trajectories for a
particle in a time-dependent spherical trap for different values of
wall velocity: a) $\alpha = -0.5 \alpha_{01}$, b) $\alpha = -2
\alpha_{01}$, c) $\alpha = 0.5 \alpha_{01}$ and d) $\alpha = 2
\alpha_{01}$. Particle is initially in the energy eigenstate
$u_{010}(r, \theta, \phi)$ and locates on $x$-axis. Red curve shows
the position of the wall.}
\vspace*{0.1cm}
\label{fig: traj_spherical_1D}
\end{figure}

Figure \ref{fig: traj_spherical_1D} represents trajectories for
particle in a spherical trap. Here, particle initially is in the
ground state $u_{010}(r, \theta, \phi) = \sqrt{2/a} \sin(\pi r/a)
Y_{00}(\Omega)/r$ which corresponds to a particle in a 1D square
box \cite{Mo-EPL-2012}.

So far we have studied a single-particle system. Now, we consider a
two-body system which are confined in our time-dependent spherical
trap. A novel feature of the causal description of a many-body
system is the non-local connection of particles: dependence of
instantaneous motion of any one particle to the coordinates of all
other particles at the same time. In this case the particle
trajectories ${\bf{r}}_i(t)$ are determined by the following system
of two simultaneous differential equations
\begin{eqnarray} \label{eq: two_par_traj}
\frac{ d{\bf{r}}_i(t) }{d t} &=& \frac{\hbar}{\mu_i}~ {\text{Im}} \left(
{\frac{\vec{\nabla}_i \Psi({\bf{r}}_1, {\bf{r}}_2, t)}
{\Psi({\bf{r}}_1, {\bf{r}}_2, t)}} \right) \bigg|_{ {\bf{r}}_j =
{\bf{r}}_j(t) }~,~~~~~~~~i, j = 1, 2.
\end{eqnarray}
If the particles are identical the wavefunction of the system must
be symmetrized. There are different statistics, Fermi-Dirac (FD) for
which total wavefunction must be antisymmetric with respect to the
exchange of particles in the system and Bose-Einstein (BE) for which
total wavefunction is symmetric under the exchange of particles.
When particles are distinguishable they are independent obeying
Maxwell-Boltzmann (MB) statistics. In Bohmian perspective the
particles are always distinguishable in all cases
\cite{Ho_book_1993}. The initial wavefunction is taken to be
\begin{eqnarray} \label{eq: twopar_initial_wave}
\Psi_{\pm}({\bf{r}}_1, {\bf{r}}_2, 0) &=& \frac{1}{ \sqrt{2} }
\left[ u_{010}({\bf{r}}_1, 0) u_{020}({\bf{r}}_2, 0) \pm
u_{020}({\bf{r}}_1, 0) u_{010}({\bf{r}}_2, 0) \right]
\nonumber \\
&=& \frac{ \sqrt{2} }{a} \frac{1}{r_1 r_2} \left[ \sin \left(
\frac{\pi r_1}{a} \right) \sin \left( \frac{2 \pi r_2}{a} \right)
\pm \sin \left( \frac{2 \pi r_1}{a} \right) \sin \left( \frac{\pi
r_2}{a} \right) \right] Y_{00}(\Omega_1) Y_{00}(\Omega_2)~,
\end{eqnarray}
which is independent of polar and azimuthal angles and so the motion
is one-dimensional. As particles are classically non-interacting
this expression will be preserved by the Schr\"{o}dinger evolution.

\begin{figure}
\centering
\includegraphics[width=12cm,angle=0]{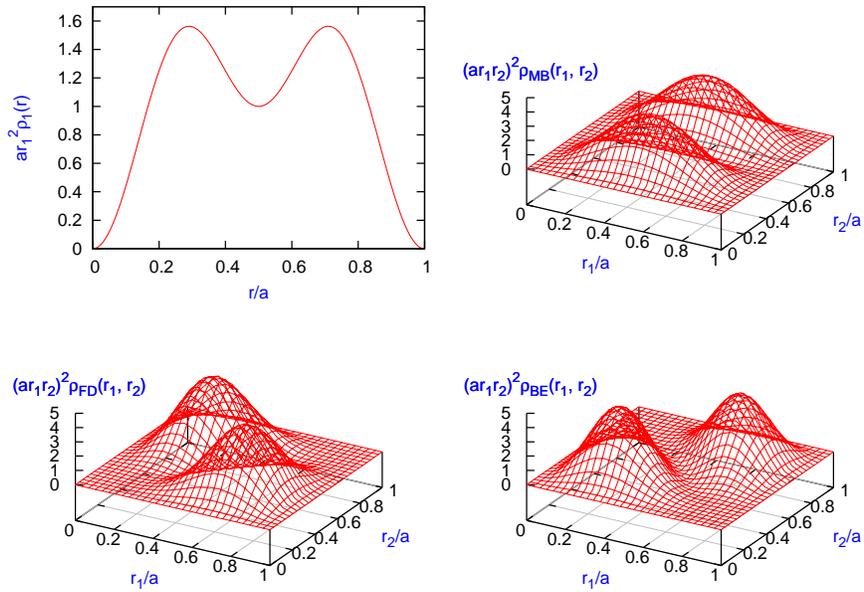}
\caption{(Color online) Single-particle and two-particle initial
distribution functions $\rho_1(r, 0)$ and $\rho_2(r_1, r_2, 0)$ for
a two-particle system confined in a time-dependent spherical trap.}
\vspace*{0.1cm}
\label{fig: rho_01}
\end{figure}

\begin{figure}
\centering
\includegraphics[width=12cm,angle=-90]{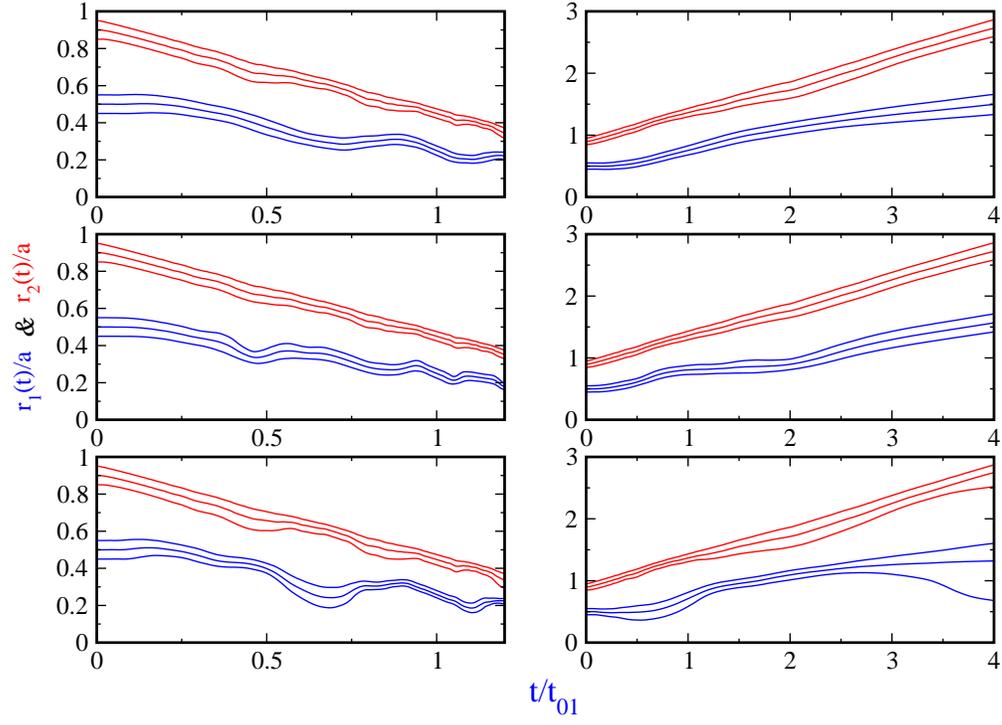}
\caption{(Color online) A selection of two-particle trajectories
$r_1(t)$ and $r_2(t)$: Maxwell-Boltzmann (first row), Fermi-Dirac
(second row) and Bose-Einstein (third row). In the left column box
is in contraction with rate $\alpha = -0.5 \alpha_{01}$ while in the
right one it expands with rate $\alpha = 0.5 \alpha_{01}$.}
\vspace*{0.5cm}
\label{fig: twoparticle_traj}
\end{figure}

\begin{figure}
\centering
\includegraphics[width=12cm,angle=-90]{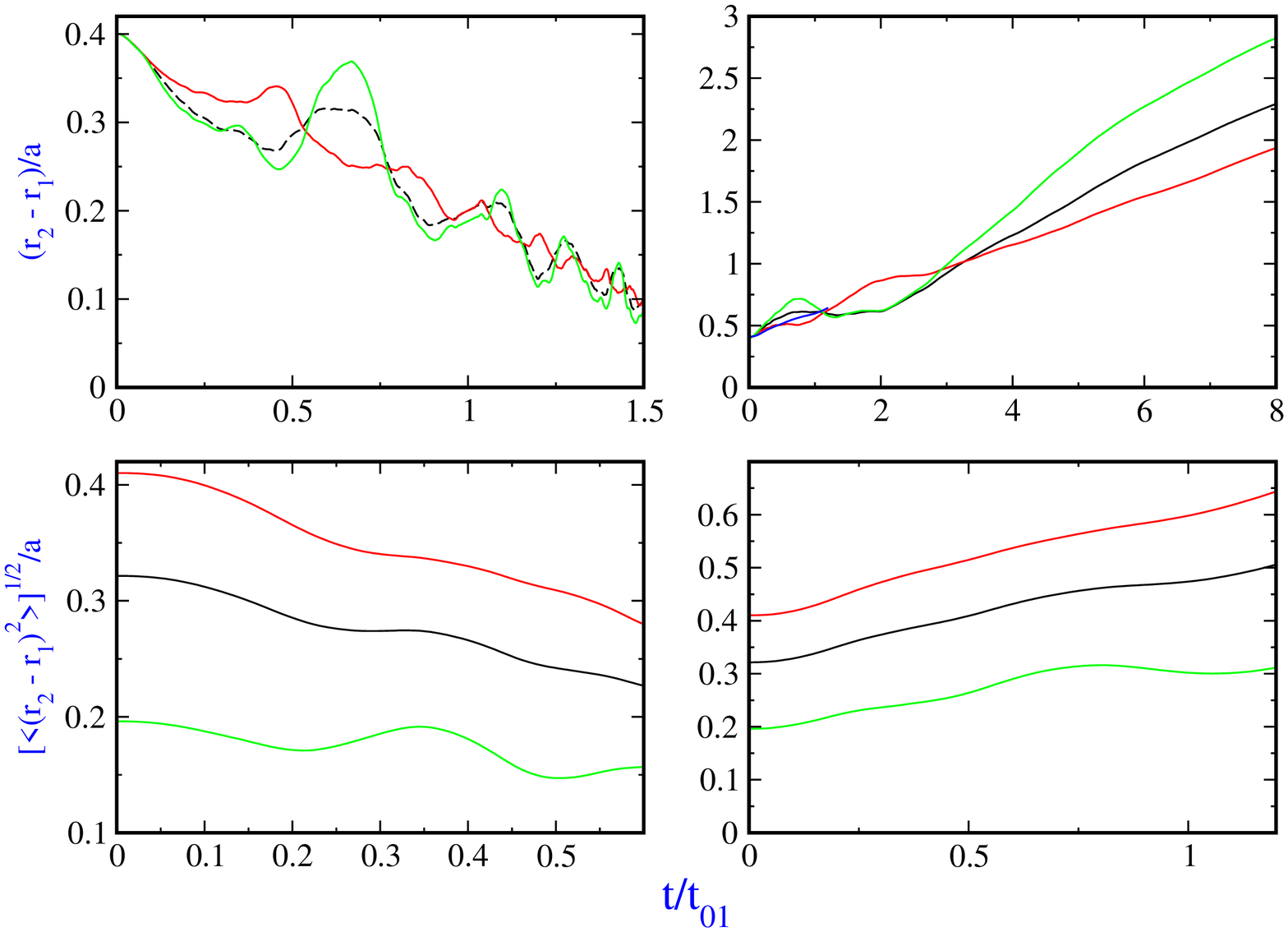}
\caption{(Color online) The relative particle separation $r_2(t) -
r_1(t)$ (first row) and root mean square separation $\sqrt{ \langle
(r_2 - r_1)^2 \rangle(t)}$ (second row) for two particles confined
in a time-dependent spherical trap obeying  Maxwell-Boltzmann
statistic (black curves), Fermi-Dirac (red curves) and Bose-Einstein
(green curves). In the left column wall velocity is $\alpha = -0.5
\alpha_{01}$  and in the right one it is $\alpha = 0.5 \alpha_{01}$.}
\vspace*{0.1cm}
\label{fig: rms_sep}
\end{figure}

An interesting quantity which shows density probability for particle 1
being in ${\bf{r}}$ regardless of the position
of the particle 2, is $ \rho_1({\bf{r}}) = \int d\Omega_2 \int_0^a dr_2
r_2^2 |\Psi_{\pm}({\bf{r}}, {\bf{r}}_2)|^2$. Initial one-particle
distribution function $ \rho_1(r) $ and two-particle density $
\rho({\bf{r}}_1, {\bf{r}}_2) = |\Psi_{\pm}({\bf{r}}_1,
{\bf{r}}_2)|^2 $ have been depicted in figure \ref{fig: rho_01} for
three different statistics. Figure \ref{fig: twoparticle_traj}
displays a selection of two-particle trajectories $r_1(t)$ and
$r_2(t)$ with initial positions $r_2(0) = r_1(0)+0.4 a$. Distance
between two particles has been shown in figure \ref {fig: rms_sep}
for a given initial conditions. This figure shows that depending on
the initial conditions relative motions of particles are determined.
There are situations for which distance between bosons is greater
than the fermions. On the other hand average separation of particles
confirms one's expectations. The mean separation $ \langle r_2 - r_1
\rangle$ is zero for fermions and bosons due to the symmetry of the
wavefunction under exchange of particles, while the root mean square
separation is less (more) in BE (FD) case than in the MB one, but
these results do not reveal details of individual motions.
Similar result has been reported for a two-particle system
composed of two identical and independent 1D harmonic oscillators
where the one-particle wavefunctions taken to be nondispersive packets
oscillating between two symmetrical points with respect to the origin \cite{Ho_book_1993}.
It's worth mentioning that in the case of a non-moving box
evolved wavefunction is given by
\begin{eqnarray*}
\Psi_{\pm}({\bf{r}}_1, {\bf{r}}_2, t) &=& \frac{1}{ \sqrt{2} }
\left[ u_{010}({\bf{r}}_1, 0) u_{020}({\bf{r}}_2, 0) \pm
u_{020}({\bf{r}}_1, 0) u_{010}({\bf{r}}_2, 0) \right] e^{-i(E_{01}-E_{02})t/\hbar}; ~~~~~~E_{ln} = \frac{\hbar^2}{2\mu a^2} x_{ln}^2~.
\end{eqnarray*}
Thus, both particles stay at rest. Moving wall makes the system
non-stationary as a result of which Bohmian particles move.


\section{Summary and Discussion} \label{sec: summary}

In this work the problem of a one-body and a two-body system
confined in time-dependent traps were studied for particular initial conditions. For the
one-body case we gave analytical relations for quantum effective
force. It was seen that when the particle is initially in an energy
eigenstate of a smaller box, irrespective of the size of the smaller
box expectation value of the momentum
operator does not change with time while for the corresponding 1D
system the behavior is different.
A selection of Bohmian trajectories computed for a
particle initially in an energy eigenstate of the time-dependent circular
and spherical boxes for different values of the wall velocity.

For the two-particle system details of motions implied by nonfactorizable wavefunctions are
subtle: one cannot say that fermions are repelled and bosons are
attracted, in comparison with distinguishable particles obeying MB statistics.
The root mean square separation confirms one's expectation:
it is less (more) in BE (FD) case than in the MB one. In both one-body and two-body system moving wall
makes the system to be non-stationary as a result of which Bohmian particles move.

{\bf{Acknowledgment}}
Financial support of the University of Qom is acknowledged.


\pagebreak


\begin{thebibliography} {99}
%
\bibitem{1D_Box}
S. W. Doescher and M. H. Rice, {\it Am. J. Phys.} {\bf 37} (1969) 1246; \\
D. W. Schlitt and C. Stutz, {\it Am. J. Phys.} {\bf 38} (1970) 70; \\
D. N. Pinder, {\it Am. J. Phys.} {\bf 58} (1990) 54; \\
M. G. da Luz and Bin Kang Cheng, {\it J. Phys. A} {\bf 25} (1992) L1043; \\
V. V. Dodonov, A. B. Klimov and D. E. Nikonov, {\it J. Math. Phys.} {\bf 34}, (1993) 3391; \\
O. Fojon, M. Gadella, L.P. Lara, {\it Comput. Math. Appl.} {\bf 59}
(2010) 964
%
\bibitem{Mo-PS-2012}
S. V. Mousavi, {\it Phys. Scr.} {\bf 86} (2012) 035004
%
\bibitem{DoAn-PLA-2000-2002}
V. V. Dodonov and M. A. Andreata, {\it Phys. Lett. A} {\bf 275}, (2000) 173;\\
V. V. Dodonov and M. A. Andreata, {\it Laser Phys.} {\bf 12}, (2002) 57
%
\bibitem{Yu-PLA-2004}
Y\"uce C., {\it Phys. Lett. A} {\bf 327} (2004) 107 \\
S. V. Mousavi, {\it Phys. Lett. A}, {\bf 377} (2013) 1513
\bibitem{Mo-EPL-2012}
S. V. Mousavi, {\it EuroPhys. Lett.}, {\bf 99} (2012) 30002
%
%
\bibitem{BoI&II_PR_1952}
D. Bohm {\it Phys. Rev} {\bf 85} (1952) 166; \\
D. Bohm {\it Phys. Rev} {\bf 85} (1952) 180
%
\bibitem{Ho_book_1993}
P. R. Holland 1993, {\it The Quantum Theory of Motion} (Cambridge: Cambridge University Press), pp. 277-310
%
%
\end{thebibliography}
\end{document}